\documentclass[fp,twocolumn]{jpsj3}
\usepackage{txfonts,color}

\title{Effects of Antisite Defects on Seebeck Coefficient in Fe$_2$VAl \\
- Analyses based on Bipolar Random Anderson Model}

\author{Takami Tohyama$^1$\thanks{tohyama@rs.tus.ac.jp} and Hidetoshi Fukuyama$^2$}
\inst{$^1$Department of Applied Physics, Tokyo University of Science, Katsushika, Tokyo 125-8585, Japan \\
$^2$Research Institute for Science and Technology, Tokyo University of Science, Shinjuku, Tokyo 162-8601, Japan} 

\abst{
A microscopic mechanism is proposed for a dramatic sign change of the Seebeck coefficient from positive to negative sign by the introduction of antisite defects in Fe$_2$VAl based on bipolar random Anderson model (BPRAM), which incorporates hybridization effects between randomly distributed antisites and host bands, where the valence and conduction bands are treated separately due to their separation in momentum space. Applying a self-consistent T-matrix approximation, we find that antisite defects in Fe$_2$VAl induce new states in the band overlap region, resulting in a scattering rate that is higher for hole carriers in the valence band than that for electron carriers in the conduction band, leading to negative Seebeck coefficient. This mechanism of sign change presents a potential new approach for controlling thermoelectric properties in semimetallic systems without changing carrier concentration.}

\begin{document}
\maketitle

\section{Introduction}
\label{Sec1}
The Heusler-type intermetallic Fe$_2$VAl compound is a nonmagnetic semimetal with a pseudogap at the Fermi level $\varepsilon_\mathrm{F}$ and exhibits a semiconductor-like resistivity behavior over a wide temperature range~\cite{Nishino1997,Okamura2000,Soda2005}. First-principles calculations have predicted a semimetallic band structure with the valence-band top at the $\Gamma$ point and the conduction-band bottom at the X point with small band overlap leading to a pseudogap at $\varepsilon_\mathrm{F}$~\cite{Guo1998,Singh1998,Weht1998,Weinert1998,Rai2017}. This particular electronic state of Fe$_2$VAl is the source of variety of electronic properties introduced by alloying and/or quenching processes, which make this system to attract a lot of attention as potential candidates for thermoelectric materials~\cite{Nishino2011}. In stoichiometric Fe$_2$VAl, the Seebeck coefficient $S$ exhibits positive values over a wide temperature range~\cite{Nishino2001},indicating that hole carriers are the majority. Notably, the sign and magnitude of $S$ are highly sensitive to off-stoichiometry, which arises from small deviations in composition ratios~\cite{Nishino2001,Hanada2001,Lue2002,Nakama2005,Miyazaki2014,Alleno2023}. This sensitivity primarily stems from minor changes in carrier concentration that shift $\varepsilon_\mathrm{F}$ within the pseudogap. Actually, even in the stoichiometric case, negative $S$ has been reported~\cite{Nakama2005,Kurosaki2009,Sk2018,Sk2024}.

Recently, it has been indicated that the introduction of antisite defects in Fe$_2$VAl through thermal quenching, but keeping stoichiometric composition, changes $S$ to negative values~\cite{Garmroudi2022}. Simultaneously, the semiconductor-like resistivity in pristine Fe$_2$VAl changes to a metallic behavior at low temperatures. These remarkable changes resemble those seen in off-stoichiometric Fe$_2$V$_{1+x}$Al$_{1-x}$ with $x>0$~\cite{Miyazaki2014}, yet their underlying mechanism must differ, since compositional change does not occur through thermal quenching.

In this paper, we propose a microscopic mechanism for the sign change of $S$ induced by antisite defects in stoichiometric Fe$_2$VAl. We consider an antisite V on the Fe sublattice, denoted as V$_\mathrm{Fe}$, and an antisite Fe on the V sublattice, denoted as Fe$_\mathrm{V}$, as impurities, which are described by the Anderson model~\cite{Anderson1961}. Since antisites in Fe$_2$VAl are randomly distributed with macroscopic concentration, we extend the single-impurity Anderson model to a system with randomly distributed impurities, referring to it as the random Anderson model (RAM).  Considering that (1) the valence band maximum and conduction band minimum are located at different points in the Brillouin zone, i.e., the $\Gamma$ point for the former and the X point for the latter and (2) the valence and conduction bands are predominantly composed of Fe$3d$ and V$3d$ states, respectively, we treat V$_\mathrm{Fe}$ and Fe$_\mathrm{V}$ as antisites in distinct bands. Furthermore, bipolar effects~\cite{Delves1965} due to the existence of both valence and conduction bands are taken into account through consideration of the fact that total number of electrons are conserved during thermal quenching in stoichiometric Fe$_2$VAl. Applying a self-consistent T-matrix approximation to this bipolar random Anderson model (BPRAM), we find that the scattering rate of holes in the valence band, induced by V$_\mathrm{Fe}$ antisites, is higher than that of electrons in the conduction band. This results in a reduced contribution of hole carriers to $S$, ultimately causing a sign change of $S$ from positive values in pristine Fe$_2$VAl  to negative values in quenched Fe$_2$VAl. This sign-change mechanism differs from the one associated with half-filling of the narrow impurity band above the valence band~\cite{Garmroudi2022}, which is one-band effect. Actually, there is no hole-electron sign change within the impurity band above the valence band, since all states result from the valence band. Such is also the case in impurity band at the bottom of conduction as had been indicated explicitly~\cite{Fukuyama1970}. In the present mechanism, the difference in scattering rates between two carriers plays a crucial role in determining the sign of $S$. This presents a potential new approach for controlling thermoelectric properties in semimetallic systems without changing carrier concentration.

This paper is organized as follows.  Section~\ref{Sec2} introduces the BPRAM alongside the methodology for calculating thermoelectric properties. In Sect.~\ref{Sec3}, we present the calculated results for the density of states, spectral conductivity, resistivity, and Seebeck coefficient in pristine and quenched Fe$_2$VAl, highlighting the emergence of the sign change in $S$ due to the presence of antisites. The microscopic mechanism underlying this sign change is analyzed within the framework of BPRAM. Finally, Section~\ref{Sec4} presents a discussion of the implications of the proposed mechanism, followed by a summary.

\section{Model and method}
\label{Sec2}
\subsection{Electronic states and model Hamiltonian}
\label{Sec2-1}
\begin{figure}[tb]
\center{
\includegraphics[width=0.4\textwidth]{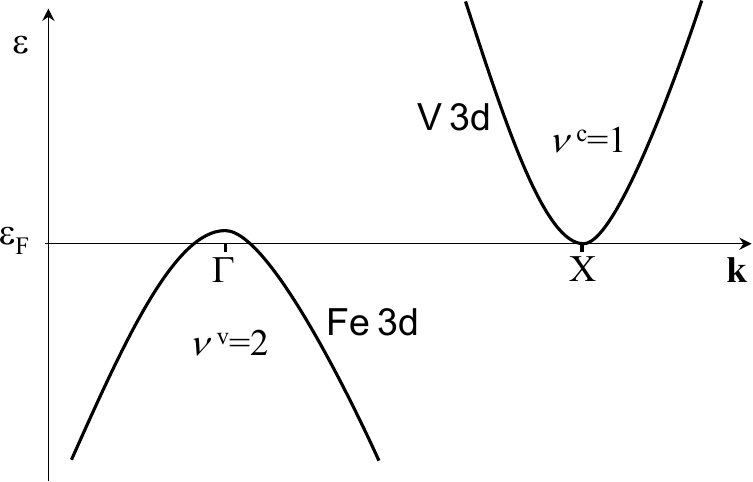}
}
\caption{Schematic illustration of the electronic states in Fe$_2$VAl. The top of the valence band at the $\Gamma$ point lies higher in energy than the bottom of the conduction band at the X point, resulting in a negative band gap ($E_\mathrm{g} < 0$). The valence band is predominantly composed of Fe$3d$ orbitals with a degeneracy of $\nu^\mathrm{v} = 2$, while the conduction band mainly consists of V$3d$ orbitals with a degeneracy of $\nu^\mathrm{c} = 1$. To reproduce the experimental Seebeck coefficient (see Fig.~\ref{fig3}(c)), the Fermi energy $\varepsilon_\mathrm{F}$ is positioned near the top of the valence band.}
\label{fig1}
\end{figure}

Electronic structure calculations using density functional theory (DFT), based on local density or generalized gradient functional approximations, have shown that Fe$_2$VAl exhibits semimetallic behavior with a negative gap $E_\mathrm{g}<0$, meaning that the valence-band top at $\Gamma$ is higher in energy than the conduction-band bottom at X~\cite{Guo1998,Singh1998,Weht1998,Weinert1998,Rai2017}.  To improve agreement with experimental transport properties of Fe$_2$VAl-based compounds, hybrid functionals and on-site electron correlation have been incorporated into DFT~\cite{Bilc2011,Do2011,Kristanovski2017,Naydenov2020,Hinterleitner2020}. Under these approaches, the gap is found to be either slightly positive~\cite{Bilc2011,Do2011,Kristanovski2017,Naydenov2020} or negative~\cite{Hinterleitner2020}. Based on these DFT calculations, simple models with both the valence and conduction bands have been developed to address experimental observations of resistivity, the Seebeck coefficient, and the Hall effect~\cite{Hinterleitner2020,Anand2020,Garmroudi2023a}. Based on these studies of Fe$_2$VAl, we construct a model with a negative band gap ($E_\mathrm{g} < 0$), in which the valence band consists of two degenerate bands, while the conduction band consists of a single band~\cite{Hinterleitner2020}. Figure~\ref{fig1} illustrates this model, where the values of $E_\mathrm{g}$ and $\varepsilon_\mathrm{F}$ are determined to reproduce the experimental Seebeck coefficient (see Sect.~\ref{Sec3}).

The effects of antisite defects on the electronic structure of Fe$_2$VAl have been investigated using DFT~\cite{Bilc2011,Venkatesh2013,Bandaru2017,Berche2020}. While V$_\mathrm{Al}$ antisites increase $E_\mathrm{g}$~\cite{Berche2020}, Fe$_\mathrm{V}$ and V$_\mathrm{Fe}$ antisites induce spin-split sharp peaks near $\varepsilon_\mathrm{F}$ in the density of states (DOS)~\cite{Garmroudi2022,Bandaru2017}. This suggests that Fe$_\mathrm{V}$ and V$_\mathrm{Fe}$ antisites significantly modify transport properties by interacting with underlying electronic states in the valence V3$d$ and conduction Fe3$d$ band, respectively. A Fe$_\mathrm{V}$ (V$_\mathrm{Fe}$) antisite can be treated as an impurity in the valence (conduction) band, which can be effectively described by the resonance state of the Anderson model~\cite{Anderson1961}. However, since Fe$_\mathrm{V}$ and V$_\mathrm{Fe}$ antisites are randomly distributed within the host sublattice, a suitable model for describing their effects is an Anderson-type model with randomly distributed resonance state impurities, referred to the RAM. Additionally, charge transfer between an antisite defect and a host band may occur. Assuming that Fe$_\mathrm{V}$ and V$_\mathrm{Fe}$ exist in equal concentrations in stoichiometric Fe$_2$VAl, we keep charge conservation, where the number of electrons transferred from Fe$_\mathrm{V}$ to the conduction band must equal the number of holes transferred from V$_\mathrm{Fe}$ to the valence band. This charge redistribution may give rise to bipolar effects in Fe$_2$VAl. To account for this phenomenon, we introduce the RAM with bipolarity, referred to the BPRAM. 

Since the valence-band top and conduction-band bottom are located at different momenta, we assume no direct interplay between the two bands except the conservation of total number of electrons. Therefore, we define the Hamiltonian of the RAM for the valence (v) and conduction (c) band separately as 
\begin{eqnarray}
H^\alpha_\mathrm{RAM}&=& \sum_{\mathbf{k},s} \varepsilon^\alpha_\mathbf{k} c^{\alpha\dagger}_{\mathbf{k}s} c^\alpha_{\mathbf{k}s} + \sum_i \left(\varepsilon^\alpha_\mathrm{d} n^\alpha_i + U^\alpha_\mathrm{d} n^\alpha_{i\uparrow}n^\alpha_{i\downarrow}\right) \nonumber \\
 &+&V^\alpha \sum_{\mathbf{k},s,i} \left( d^{\alpha\dagger}_{is} c^\alpha_{\mathbf{k}s} + \mathrm{H.c.}  \right),
\label{HRA}
\end{eqnarray}
where $\alpha=$v or $\alpha=$c, $\varepsilon^\alpha_\mathbf{k}$ is a momentum $\mathbf{k}$-dependent dispersion, $\varepsilon^\alpha_\mathrm{d}$ is the energy level of antisite defects, $U^\alpha_\mathrm{d}$ is on-site Coulomb interaction on impurities, $V^\alpha$ is a hybridization parameter between mobile and impurity electrons. The operator $c^{\alpha\dagger}_{\mathbf{k}s}$ creates a mobile electron with $\mathbf{k}$ and spin $s$, the operator $d^{\alpha\dagger}_{is}$ creates a localized electron on an impurity site $i$, and $n^\alpha_i=n^\alpha_{i\uparrow}+n^\alpha_{i\downarrow}$ with $n^\alpha_{is}=d^{\alpha\dagger}_{is} d^\alpha_{is}$.

\subsection{Self-consistent T-matrix approximation}
\label{Sec2-2}
Electrons in the valence and conduction bands are scattered by random impurities via a $\mathbf{k}$-independent hybridization term in Eq.~(\ref{HRA}). This effect is characterized by a $\mathbf{k}$-independent but spin-dependent self-energy for mobile electrons, $\Sigma_s^\alpha(\varepsilon)$. Following the original study by Anderson~\cite{Anderson1961}, the Green's function for the valence and conduction electrons is given by
\begin{eqnarray}
G_s^\alpha(\mathbf{k},\varepsilon)=\frac{1}{\varepsilon-\varepsilon^\alpha_\mathbf{k} -\Sigma_s^\alpha(\varepsilon)}.
\label{Gvc}
\end{eqnarray}
The Green's function for impurities is assumed to be independent of site $i$, given by
\begin{eqnarray}
G_s^{\mathrm{d},\alpha}(\varepsilon)=\frac{1}{\varepsilon-\varepsilon^{\mathrm{d},\alpha}_s-\Sigma_s^{\mathrm{d},\alpha}(\varepsilon)},
\label{Gl}
\end{eqnarray}
where $\varepsilon_s^{\mathrm{d},\alpha}=\varepsilon^\alpha_\mathrm{d}+U^\alpha_\mathrm{d} \left<n^\alpha_{-s}\right>$ with $\left<n^\alpha_s\right>=-\frac{1}{\pi}\int_{-\infty}^{\varepsilon_\mathrm{F}}\mathrm{Im} G_s^{\mathrm{d},\alpha}(\varepsilon)d\varepsilon$. The self-energy $\Sigma_s^{\mathrm{d},\alpha}(\varepsilon)$, which is assumed to proportional to the product of $(V^\alpha)^2$ and the local Green's function of mobile electrons $g_s^\alpha(\varepsilon)=\sum_\mathbf{k} G_s^\alpha(\mathbf{k},\varepsilon)$, i.e., $\Sigma_s^{\mathrm{d},\alpha}(\varepsilon)=(V^\alpha)^2 g_s^\alpha(\varepsilon)$. 

\begin{figure}[tb]
\center{
\includegraphics[width=0.45\textwidth]{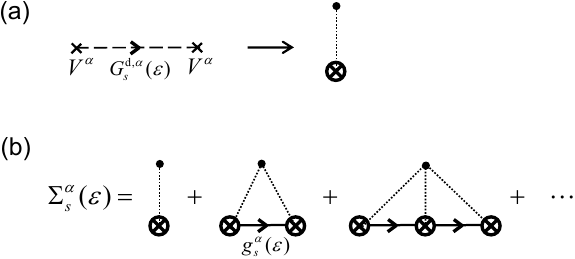}
}
\caption{The diagram of the self energy for valence ($\alpha=\mathrm{v}$) and conduction ($\alpha=\mathrm{c}$) electrons. (a) The case of a single impurity. In the left diagram, the dashed line represents the Green's function for an impurity, $G_s^{\mathrm{d},\alpha}(\varepsilon)$, while the crosses denote the hybridization interaction $V^\alpha$. We symbolize the diagram as $\mathbf{\otimes}$, combined with a dotted line with a solid circle representing an impurity site. (b) Single-site approximation for impurities and self-consistent T-matrix approximation. The solid lines denote the local Green's function, $g_s^\alpha(\varepsilon)$, for valence and conduction electrons.}
\label{fig2}
\end{figure}

For a single impurity, the self-energy $\Sigma_s^\alpha(\varepsilon)$ for valence and conduction electrons in Eq.~(\ref{Gvc}) is given by $(V^\alpha)^2 G_s^{\mathrm{d},\alpha}(\varepsilon$)~\cite{Anderson1961}, as shown in Fig.~\ref{fig2}(a). In the presence of randomly distributed impurities, we apply a single-site approximation for impurities and use a self-consistent T-matrix approximation for the self-energy $\Sigma_s^\alpha(\varepsilon)$, which is shown in Fig.~\ref{fig2}(b). This leads to a self-consistent equation for the self-energy,
\begin{eqnarray}
\Sigma_s^\alpha(\varepsilon)=c_\mathrm{i}\frac{(V^\alpha)^2G_s^{\mathrm{d},\alpha}(\varepsilon)}{1-(V^\alpha)^2G_s^{\mathrm{d},\alpha}(\varepsilon) g_s^\alpha(\varepsilon)},
\label{SE}
\end{eqnarray}
where $c_\mathrm{i}$ represents the impurity concentration. We solve Eq.~(\ref{SE}) self-consistently to obtain the self-energy for both valence and conduction electrons with common $c_\mathrm{i}$ for both conduction and valence bands.

Considering that the energy dispersion $\varepsilon^\alpha_\mathbf{k}$ near the valence-band top and conduction-band bottom in Fe$_2$VAl is well approximated in terms of effective masses~\cite{Hinterleitner2020,Anand2020,Garmroudi2023a}, we assume that the DOS per spin is given by $(1/\pi W^\alpha)\left[1-(\tilde{\varepsilon}/W^\alpha)^2\right]^{1/2}$ with half the bandwidth $W^\alpha$~\cite{Fukuyama1970}, where $\tilde{\varepsilon}=\varepsilon+W^\mathrm{v}$ and $\tilde{\varepsilon}=\varepsilon-W^\mathrm{c}-E_\mathrm{g}$ for the valence and conduction band, respectively, taking the valence-band top as the origin of energy. The local Green's function $g_s^\alpha(\varepsilon)$ corresponding to this assumption is given by~\cite{Fukuyama1970}
\begin{eqnarray}
g_s^\alpha(\varepsilon)=\frac{\nu^\alpha}{W^\alpha}\left[z_s^\alpha+i\tau_s^\alpha(1+z_s^\alpha)\right]
\label{gs}
\end{eqnarray}
with $\tau_s^\alpha=[(1-z_s^\alpha)/(1+z_s^\alpha)]^{1/2}$ and $z_s^\alpha=\left[\tilde{z}+\Sigma_s^\alpha(\tilde{z})\right]/W^\alpha$, where $\tilde{z}=\tilde{\varepsilon}+i\delta$, introducing an intrinsic damping factor $\delta$ for a system without antisites. $\nu^\alpha$ in Eq.~(\ref{gs}) represents the degeneracy of the bands and we take $\nu^\mathrm{v}=2$ and $\nu^\mathrm{c}=1$ for Fe$_2$VAl~\cite{Hinterleitner2020}. Since the valence-band mass is larger than the conduction-band mass~\cite{Hinterleitner2020}, we assume $W^\mathrm{v}<W^\mathrm{c}$. The DOS per spin is given by $D_s^\alpha(\varepsilon)=-\frac{1}{\pi}\mathrm{Im} g_s^\alpha(\varepsilon)$.

\subsection{Thermoelectric properties}
\label{Sec2-3}
In the present BPRAM, electrons in the valence and conduction bands contribute to electrical and heat currents separately. The spectral conductivity, $\sigma_s^\alpha(\varepsilon)$, which governs these current, is given by~\cite{Fukuyama1970}
\begin{eqnarray}
\sigma_s^\alpha(\varepsilon)=\frac{e^2\pi}{3}\left(\frac{F^\alpha}{W^\alpha}\right)^2L_s^\alpha(\tilde{\varepsilon})
\label{sc}
\end{eqnarray}
with
\begin{eqnarray}
&&L_s^\alpha(\tilde{\varepsilon})=\left(\mathrm{Im} z_s^\alpha\right)^2 \nonumber \\
&&\ \ \ -\frac{1}{2}\mathrm{Re}\tau_s^\alpha(1+z_s^\alpha)\left(\frac{1-(z_s^\alpha)^2}{\mathrm{Im} z_s^\alpha}+3i z_s^\alpha\right),
\label{L}
\end{eqnarray}
where $e$ is the elementary charge. We introduce a factor $F^\alpha$ on the order of band width in Eq.~(\ref{sc}). This factor was originally introduced to ensure that the prefactor of $L_s^\alpha(\tilde{\varepsilon})$ matches that in the effective mass approximation near the band edges~\cite{Fukuyama1970}. In the present study, however, we treat $F^\alpha$ as an adjustable parameter to reproduce the temperature dependence of the experimental $S$ and resistivity $\rho$ in pristine Fe$_2$VAl~\cite{Garmroudi2022}.
 
Under an electric field $\mathbf{E}$ and a temperature gradient $\nabla T$, the electrical current density $\mathbf{j}$ is described within the linear response theory as $\mathbf{j}=L_{11}\mathbf{E} + L_{12}(-\nabla T)$, with $L_{11}$ and $L_{12}$ being electrical conductivity and thermoelectric conductivity, respectively~\cite{Aschcroft}. In general, $L_{12}$ has contributions from both electrons and phonons~\cite{Ogata2019}. However, the latter contributions (phonon drag) are expected to be small here, and ignored in the following. By defining the total spectral conductivity as $\sigma_s^\mathrm{t}(\varepsilon)=\sigma_s^\mathrm{v}(\tilde{\varepsilon})+\sigma_s^\mathrm{c}(\tilde{\varepsilon})$, $L_{11}$ associated with the electric current and $L_{12}$ related to both electric and heat currents are given, respectively, by
\begin{equation}
L_{11}=\int_{-\infty}^\infty \left(-\frac{\partial f(\varepsilon)}{\partial \varepsilon}\right)\sum_s\sigma_s^\mathrm{t}(\varepsilon)d\varepsilon,
\label{L11}
\end{equation}
\begin{equation}
L_{12}=-\frac{1}{e}\int_{-\infty}^\infty \left(-\frac{\partial f(\varepsilon)}{\partial \varepsilon}\right)\left(\varepsilon-\mu\right)\sum_s\sigma_s^\mathrm{t}d\varepsilon,
\label{L12}
\end{equation}
where $f(\varepsilon)$ is the Fermi distribution function, defined as $f(\varepsilon)=1/(e^{(\varepsilon-\mu)/(k_\mathbf{B}T)}+1)$ being $k_\mathrm{B}$ the Boltzmann factor, and the chemical potential $\mu$ is a temperature-dependent function $\mu=\mu(T)$ determined to satisfy the total electron density $n=\sum_s \int_{-\infty}^\infty f(\varepsilon)D_s^\mathrm{t}(\varepsilon)d\varepsilon$ in terms of the total DOS per spin $D_s^\mathrm{t}(\varepsilon)=D_s^\mathrm{v}(\varepsilon)+D_s^\mathrm{c}(\varepsilon)$. The electric resistivity $\rho$ and the Seebeck coefficient $S$ are given by $\rho=1/L_{11}$ and $S=L_{12}/L_{11}$, respectively. 

\section{Results}
\label{Sec3}
\subsection{Pristine Fe$_2$VAl}
\label{Sec3-1}
\begin{figure}[tb]
\center{
\includegraphics[width=0.4\textwidth]{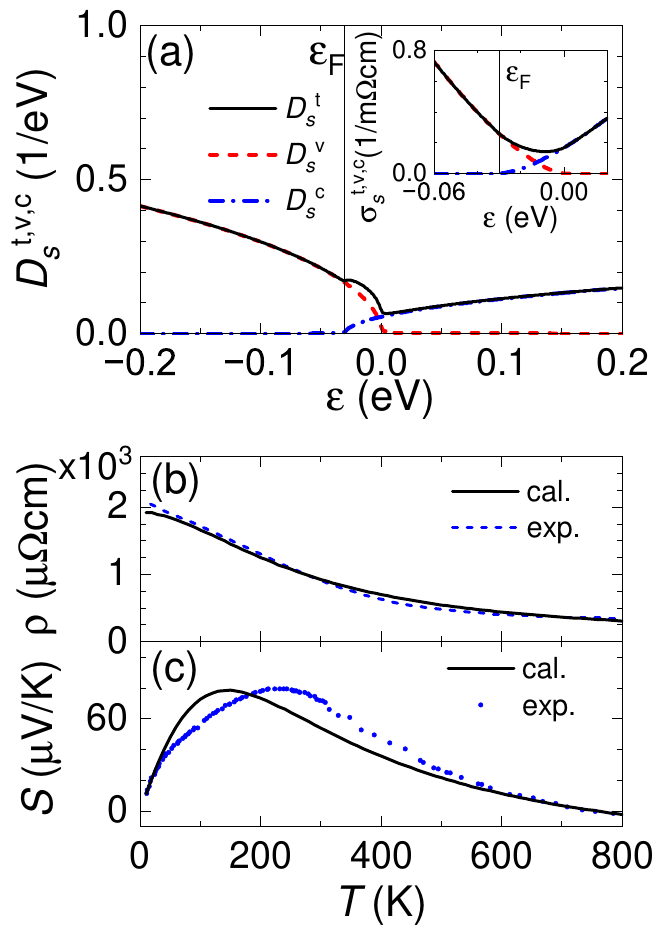}
}
\caption{(Color online) Electronic states and transport properties in Fe$_2$VAl without antisite defects. (a) The DOS per spin for total (black solid line), valence (red dashed line), and conduction (blue dashed-dotted line) electrons. The energy zero is set at the top of the valence band. Inset shows the spectral conductivity per spin for total electrons (black solid line), valence electrons (red dashed line), and conduction electrons (blue dashed-dotted line). The vertical lines denote $\varepsilon_\mathrm{F}$. (b) The temperature dependence of the calculated resistivity $\rho$ (black solid line) and experimental one (blue broken line)~\cite{Garmroudi2022}. (c) The temperature dependence of the calculated Seebeck coefficient $S$ (black solid line) and experimental one (blue dots)~\cite{Garmroudi2022}.}
\label{fig3}
\end{figure}

We regard pristine Fe$_2$VAl as a stoichiometric compound without antisite defects~\cite{Nishino1997,Nishino2001,Garmroudi2022}, where $S$ is positive over a wide temperature range of $T<750$~K with $\rho$ exhibiting a semiconducting behavior and shows a sign change around $T=750$~K. Due to the positive $S$, hole carriers are expected to dominate transport properties with $\varepsilon_\mathrm{F}$ located inside the valence band~\cite{Hinterleitner2020} probably due to small amount of off-stoichiometry. By fitting calculated temperature dependence of $S$ and $\rho$ to experimental data for pristine Fe$_2$VAl~\cite{Garmroudi2022}, we determine the following model parameters: $W^\mathrm{v}=1.52$~eV, $W^\mathrm{c}=2.0$~eV, $E_\mathrm{g}=\varepsilon_\mathrm{F}=-0.03$~eV,  $F^\mathrm{v}=7.0$~eV, $F^\mathrm{c}=11.2$~eV, and $\delta=1.0$~meV.  Figures~\ref{fig3}(b) and \ref{fig3}(c) show the calculated $\rho$ and $S$ (black solid lines), respectively, together with experimental data~\cite{Garmroudi2022} (blue broken line and dots). While the temperature showing a peak in $S$ does not perfectly align with experimental one, the sign change around $T=750$~K is well reproduced. Figure~\ref{fig3}(a) shows the DOS for the valence and conduction bands, along with the total DOS. The inset in Fig.~\ref{fig3}(a) shows the spectral conductivity, where $\sigma_s^\mathrm{t}$ (black solid line) exhibits a negative gradient near $\varepsilon_\mathrm{F}$, which is a key factor in the positive $S$ at low temperatures.

\subsection{Effects of antisites without using BPRAM  }
\label{Sec3-2}
Antisite defects of V$_\mathrm{Al}$ are known to increase the band gap $E_\mathrm{g}$~\cite{Berche2020}.  As $E_\mathrm{g}$ increases while $\varepsilon_\mathrm{F}$ remains unchanged, the contribution of conduction electrons to $S$ becomes small, as easily expected from the shift of $D_s^\mathrm{c}$ and $\sigma_s^\mathrm{c}$ toward higher energies in Fig.~\ref{fig3}(a). This leads to an enhancement of positive $S$ at low temperatures. Consequently, the observed sign change in $S$ is unlikely to be attributed to V$_\mathrm{Al}$ antisites.

In contrast, the V$_\mathrm{Fe}$ and Fe$_\mathrm{V}$ antisites directly influence the valence and conduction bands, respectively, through a hybridization term, as described by the RAM in Eq.~(\ref{HRA}). In addition,  charge transfer may occur between an antisite defect and the host band. Assuming that V$_\mathrm{Fe}$ and Fe$_\mathrm{V}$ exist in equal concentrations, $c_\mathrm{i}$, we equate the number of electrons transferred from Fe$_\mathrm{V}$ to the conduction band with the number of holes transferred from V$_\mathrm{Fe}$ to the valence band with difference of degeneracy taken into account. This semimetallic electronic state leads to a significantly negative band gap and an enhancement of the total DOS near $\varepsilon_\mathrm{F}$, resulting in a decrease in $\rho$.

Here, we use $c_\mathrm{i}=0.01$ based on an NMR study reporting the concentration of antisite defects ranging from $c_\mathrm{i}=0.003$ to $c_\mathrm{i}=0.04$~\cite{Suh2006}. For simplicity, we assume that the 1\% of DOS for the conduction band is occupied by electrons transferred from Fe$_\mathrm{V}$, with an equivalent number of holes introduced into the valence band. To satisfy this bipolar condition in both bands, we set $E_\mathrm{g}$ and $\varepsilon_\mathrm{F}$ as $E_\mathrm{g}=-0.205$~eV and $\varepsilon_\mathrm{F}=-0.075$~eV. 

\begin{figure}[tb]
\center{
\includegraphics[width=0.4\textwidth]{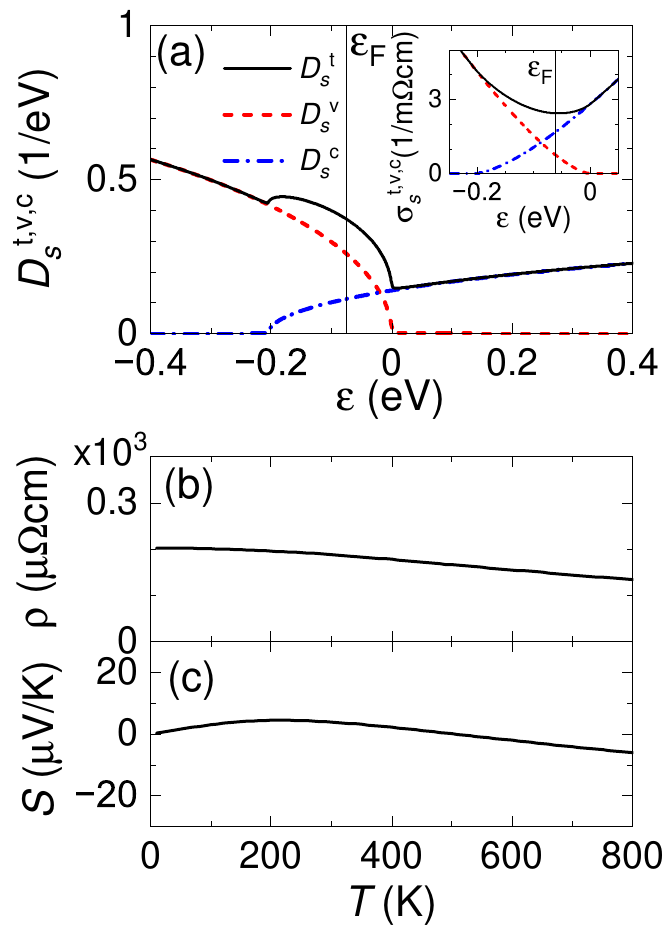}
}
\caption{(Color online) Electronic states and transport properties in Fe$_2$VAl considering only charge transfer from antisites. (a) The DOS per spin for total electrons (black solid line), valence electrons (red dashed line), and conduction electrons (blue dashed-dotted line). The energy zero is set at the top of the valence band. Inset shows the spectral conductivity per spin for total (black solid line), valence electrons (red dashed line), and conduction electrons (blue dashed-dotted line). The vertical lines denote $\varepsilon_\mathrm{F}$. (b) The temperature dependence of calculated resistivity $\rho$. (c) The temperature dependence of calculated Seebeck coefficient $S$.} 
\label{fig4}
\end{figure}

We find that maintaining charge conservation alone does not explain the sign change of $S$, as shown in Fig.~\ref{fig4}. Figure~\ref{fig4}(a) shows the resulting DOS and spectral conductivity per spin. Due to charge transfer from the antisites, the overlap between the valence and conduction bands increases compared to that in Fig.~\ref{fig3}(a), leading to an enhancement of $D_s^\mathrm{t}$ and $\rho_s^\mathrm{t}$. The resistivity $\rho$ shown in Fig.~\ref{fig4}(b) is an order of magnitudes smaller than that in the absence of antisite defects [see Fig.~\ref{fig3}(b)], consistent with experimental $\rho$ observed in quenched samples~\cite{Garmroudi2022}. However, the calculated $S$ shown in Fig.~\ref{fig4}(c) exhibits small positive values at low temperatures, which contradicts experimental $S$ showing sign change~\cite{Garmroudi2022}. Therefore, the sign change of $S$ cannot be solely attributed to charge transfer induced by antisite defects. This clearly demonstrates that the sign change of $S$ requires treating antisite defects as impurities within the BPRAM framework as will be clarified in the following.

\subsection{Effects of antisites with BPRAM}
\label{Sec3-2}

The DFT calculations have shown that V$_\mathrm{Fe}$ and Fe$_\mathrm{V}$ form localized states near $\varepsilon_\mathrm{F}$~\cite{Garmroudi2022,Bandaru2017}. Due to electron correlation, the spin degeneracy of these states is lifted, leading to spin-split states. In V$_\mathrm{Fe}$, these states are positioned across $\varepsilon_\mathrm{F}$ with a separation of approximately 0.3~eV, whereas both states in Fe$_\mathrm{V}$ lie above $\varepsilon_\mathrm{F}$. 
 Based on these observations, we set in our model the energy levels of V$_\mathrm{Fe}$ and Fe$_\mathrm{V}$ to $\varepsilon^\mathrm{v}_\mathrm{d}=-0.2$~eV and $\varepsilon^\mathrm{c}_\mathrm{d}=0.3$~eV, respectively. Assuming the same on-site Coulomb interaction for V$_\mathrm{Fe}$ and Fe$_\mathrm{V}$, i.e., $U_\mathrm{d}=U_\mathrm{d}^\mathrm{v}=U_\mathrm{d}^\mathrm{c}$, we vary $U_\mathrm{d}$ from 0~eV to 0.2~eV in the following calculations to examine its role.

\begin{figure}[tb]
\center{
\includegraphics[width=0.4\textwidth]{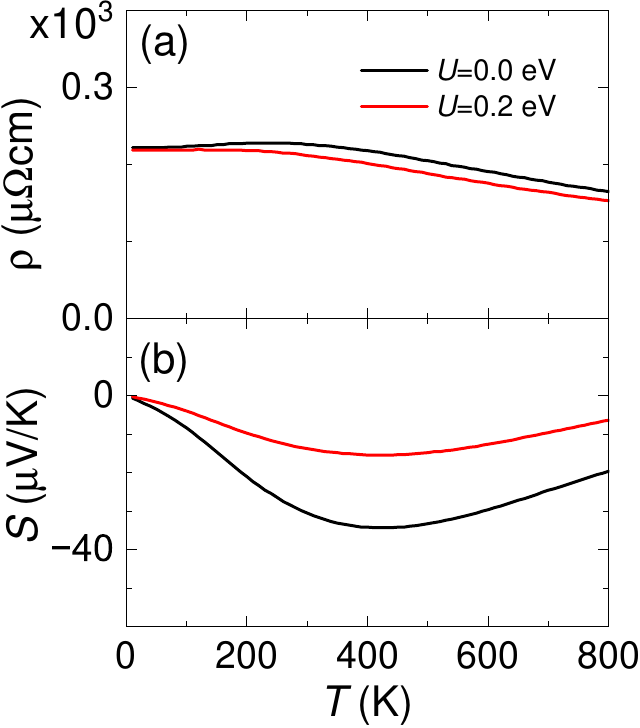}
}
\caption{(Color online) The temperature dependence of (a) calculated resistivity $\rho$ and (b) $S$ in the BPRAM describing Fe$_2$VAl with antisite defects.  Black and red solid lines represent the data for $U_\mathrm{d}=0$~eV and 0.2~eV, respectively.}
\label{fig5}
\end{figure}

Assuming hybridization parameters $V^\mathrm{v}= V^\mathrm{c}=0.1$~eV, we self-consistently determine the self-energies in Eq.~(\ref{SE}) and calculate $\rho$ and $S$ using Eqs.~(\ref{sc}), (\ref{L11}), and (\ref{L12}). The calculated values of $\rho$ and $S$ for $U_\mathrm{d}=0$~eV and 0.2~eV are shown in Figs.~\ref{fig5}(a) and \ref{fig5}(b), respectively. For both cases, the negative sign of $S$ persists up to high temperatures, indicating that the presence of on-site Coulomb interaction at antisite defects is not a critical factor in determining the sign of $S$. We note that the values of calculated $S$ align closely with experimental $S$ for quenched Fe$_2$VAl with quenching temperature of 1150${}^\circ$C~\cite{Garmroudi2022}. For $U=0$~eV, $\rho$ slightly increases with increasing $T$ at low temperatures, demonstrating metallic behavior similar to quenched Fe$_2$VAl with quenching temperature above 1000${}^\circ$C~\cite{Garmroudi2022}. As $T$ increases further, $\rho$ decreases, showing a broad peak near $T\sim 300$~K, which is lower than the temperature corresponding to the maximum of $S$. This correlation between the peak of $\rho$ and the maximum of $S$ is consistent with experimental observations in Fe$_2$VAl quenched at 1150${}^\circ$C.

\begin{figure}[tb]
\center{
\includegraphics[width=0.4\textwidth]{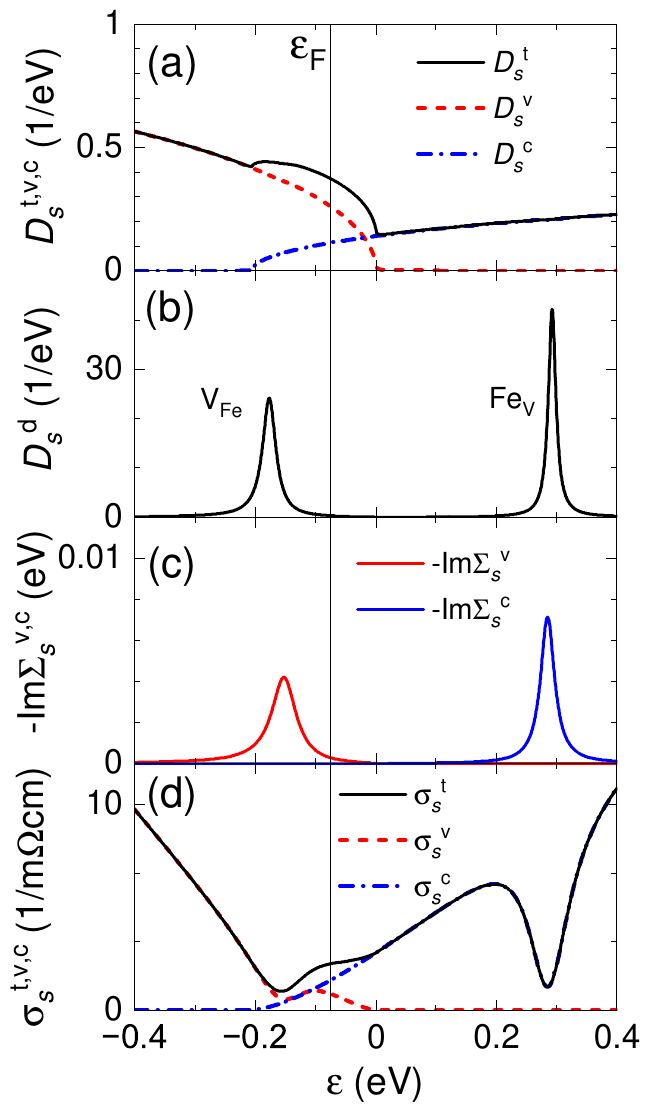}
}
\caption{(Color online) Electronic states of Fe$_2$VAl with antisite defects obtained by the BPRAM with $U_\mathrm{d}=0$~eV. (a) The DOS per spin for total electrons (black solid line), valence electrons (red dashed line), and conduction electrons (blue dashed-dotted line). (b) The DOS per spin for V$_\mathrm{Fe}$ and Fe$_\mathrm{V}$ antisites. (c) The imaginary part of spin-dependent self-energy for the valence (red line) and conduction (blue line) electrons. (d) The spectral conductivity per spin for total (black solid line), valence (red dashed line), and conduction (blue dashed-dotted line) electrons. The vertical line represent $\varepsilon_\mathrm{F}$.}
\label{fig6}
\end{figure}

To elucidate the origin of the sign change in $S$, we examine the electronic states per spin in the BPRAM with $U_\mathrm{d}=0$~eV, as shown in Fig.~\ref{fig6}. Figure~\ref{fig6}(b) presents the DOS for antisites, defined as $D_s^\mathrm{d}=-\frac{1}{\pi}\mathrm{Im}(G_s^\mathrm{d,v}+G_s^\mathrm{d,c})$, which reveals impurity states of V$_\mathrm{Fe}$ and Fe$_\mathrm{V}$ located within the valence and conduction band, respectively. Because of hybridization with antisites, the scattering rates of valence and conduction electrons represented by $-\mathrm{Im}\Sigma_s^\alpha$ increase near the impurity levels, as shown in Fig.~\ref{fig6}(c). In the DOS shown in Fig.~\ref{fig6}(a), such an enhanced scattering rate induces a subtle non-monotonic behavior at $\varepsilon=-0.16$~eV in $D_s^\mathrm{v}$. A similar feature appears in $D_s^\mathrm{c}$ at $\varepsilon=0.28$~eV, though it is barely visible. Such non-monotonic feature in the DOS is a hallmark of coupling with impurity, as reported in studies of the periodic Anderson model with random potential~\cite{Grenzebach2008}.

The influences of antisites on transport properties in the presence of large scattering rates is also evident in the spectral conductivity, shown in Fig.~\ref{fig6}(d). Here, $\sigma_s^\mathrm{c}$, equivalently $\sigma_s^\mathrm{t}$, is significantly suppressed at $\varepsilon=0.28$~eV. Likewise, $\sigma_s^\mathrm{v}$ at $\varepsilon=-0.16$~eV is suppressed due to the presence of large scattering rates at this energy. Consequently, near $\varepsilon_\mathrm{F}$, $\sigma_s^\mathrm{c}$ coming from conduction electrons largely contribute to the spectral conductivity, despite the DOS near $\varepsilon_\mathrm{F}$ being predominantly composed of valence electrons as seen in Fig.~\ref{fig6}(a). This suppression of valence-band spectral conductivity near $\varepsilon_\mathrm{F}$  significantly reduces the contribution of hole carriers to $S$, leading to negative $S$, as shown in Fig.~\ref{fig5}(b). Thus, the negative sign of $S$ can be attributed to the increased scattering rates of valence electrons caused by V$_\mathrm{Fe}$ antisites.  To further increase $S$, it is necessary to increase the scattering rate, for example, by increasing $c_\mathrm{i}$.

\begin{figure}[tb]
\center{
\includegraphics[width=0.4\textwidth]{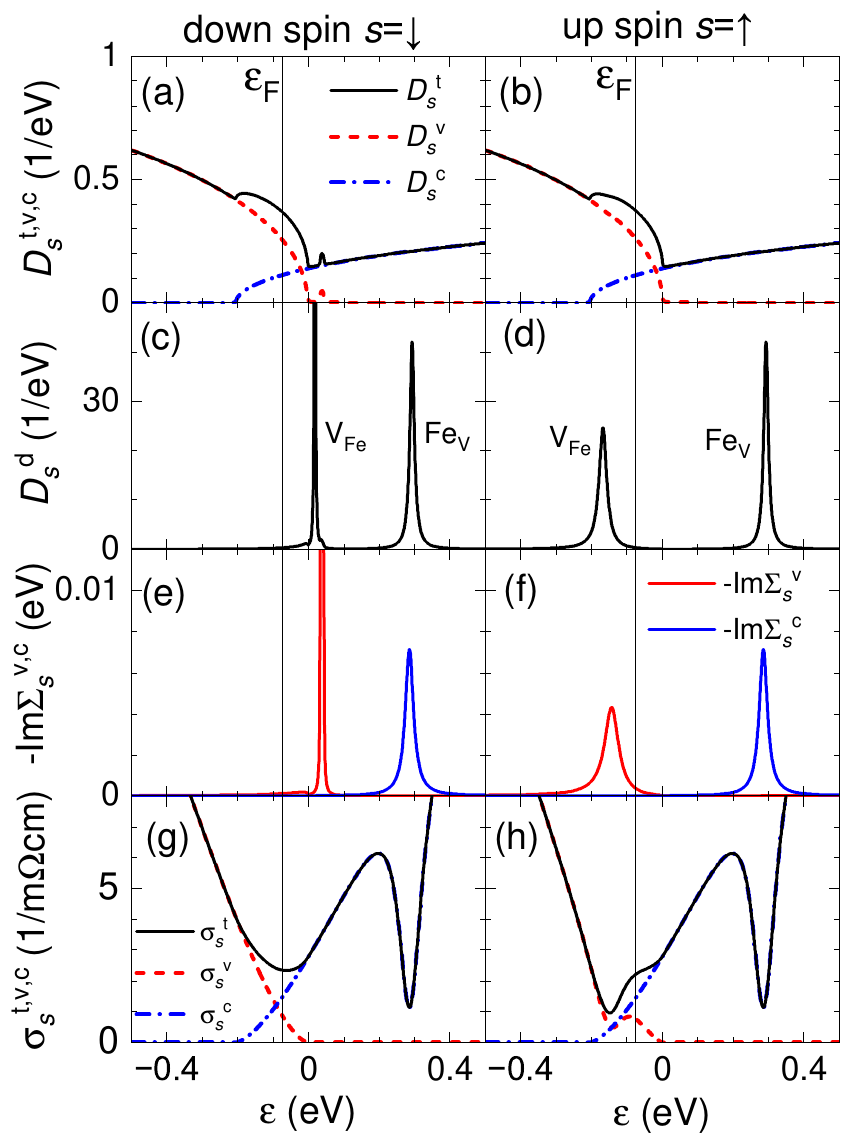}
}
\caption{(Color online) Electronic states of Fe$_2$VAl with antisite defects obtained by the BPRAM with finite $U_\mathrm{d}=0.2$~eV. (a), (b) The DOS per spin for total electrons (black solid line), valence electrons (red dashed line), and conduction electrons (blue dashed-dotted line). (c), (d) The DOS per spin for V$_\mathrm{Fe}$ and Fe$_\mathrm{V}$ antisites. (e), (f) The imaginary part of spin-dependent self-energy for the valence (red line) and conduction (blue line) electrons. (g), (h) The spectral conductivity per spin for total (black solid line), valence (red dashed line), and conduction (blue dashed-dotted line) electrons. (a), (c), (e), and (g) [(b), (d), (f), and (h)] are for down [up] spin. The vertical lines denote $\varepsilon_\mathrm{F}$. }
\label{fig7}
\end{figure}

We note that the same mechanism remains valid even when $U_\mathrm{d}$ is finite. Figure~\ref{fig7} exhibits the electronic states of the BPRAM with $U_\mathrm{d}=0.2$~eV, where the left and right columns correspond to down-spin and up-spin quantities, respectively. In Figs.~\ref{fig7}(c) and \ref{fig7}(d), the energy levels of V$_\mathrm{Fe}$ for down and up spins appear across $\varepsilon_\mathrm{F}$ in the impurity DOS, $D_s^\mathrm{d}$, with a separation approximately equal to $U_\mathrm{d}$. In contrast,  Fe$_\mathrm{V}$ does not exhibit spin polarization. Because of hybridization with impurities, the scattering rates of valence and conduction electrons represented by $-\mathrm{Im}\Sigma_s^\mathrm{v(c)}$ increase at the impurity level, as shown in Figs.~\ref{fig7}(e) and \ref{fig7}(f). In the DOS presented in Figs.~\ref{fig7}(a) and \ref{fig7}(b), the increased scattering rate induces a non-monotonic behavior, for example, at $\varepsilon=-0.24$~eV in the up-spin valence band.  We note that a split-off state characterized by a peak structure at $\varepsilon=0.04$~eV in $D_\downarrow^\mathrm{v}$ emerges due to an impurity level being positioned outside the band edge. This fact points to an important fact that formation of magnetic moments depends on the very subtle features of antisites.

The impact of impurities, arising from large scattering rates, is also evident in the spectral conductivity shown in Figs.~\ref{fig7}(g) and \ref{fig7}(h), where $\sigma_s^\mathrm{c}$, equivalently $\sigma_s^\mathrm{t}$, is significantly suppressed at $\varepsilon=0.29$~eV. Additionally, $\sigma_\uparrow^\mathrm{v}$ at $\varepsilon=-0.15$~eV is suppressed due to the presence of strong scattering rates at the same energy in the valence band, as shown in Fig.~\ref{fig7}(f).  As a result, the spectral conductivity near $\varepsilon_\mathrm{F}$ is primarily dominated by $\sigma_\uparrow^\mathrm{c}$ from conduction electrons, despite the fact that the DOS near $\varepsilon_\mathrm{F}$ is largely contributed by valence electrons, as shown in Fig.~\ref{fig7}(b). This suppression of the valence-band spectral conductivity inevitably reduces the contribution of hole carriers to $S$, resulting in negative $S$ [see Fig.~\ref{fig5}(b)], which is predominantly governed by conduction electrons. Thus, as in the  the case of $U_\mathrm{d}=0$~eV, the sign change of $S$ is again attributed to the increased scattering rates of valence electrons caused by V$_\mathrm{Fe}$ antisites. 

\section{Summary and discussions}
\label{Sec4}
\begin{figure}[tb]
\center{
\includegraphics[width=0.4\textwidth]{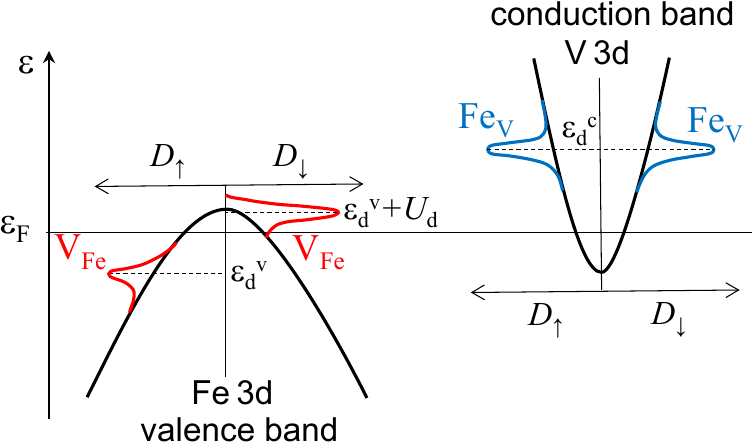}
}
\caption{(Color online) Schematic illustration of the electronic states in Fe$_2$VAl with antisites. The DOS  per spin ($D_\uparrow$ and $D_\downarrow$) for the valence band, primarily composed of Fe$3d$ orbitals (left), and the conduction band, primarily composed of V$3d$ orbitals,  is shown by the black solid lines . The $\varepsilon_\mathrm{F}$ is located near the top of the valence band and near the bottom of the conduction band. Resonance states associated with V$_\mathrm{Fe}$ and Fe$_\mathrm{V}$ are denoted by red and blue solid lines, respectively, superimposed on the DOS. The Fe$_\mathrm{V}$ resonance states appear at $\varepsilon_\mathrm{d}^\mathrm{c}$ for both spin channels, while the V$_\mathrm{Fe}$ resonance states are split by an energy $U_\mathrm{d}$ above $\varepsilon_\mathrm{d}^\mathrm{v}$ due to on-site Coulomb interaction.}
\label{fig8}
\end{figure}

As demonstrated in Sect.~\ref{Sec3-2}, the sign change of $S$ from positive value in pristine Fe$_2$VAl to negative value in Fe$_2$VAl with antisite defects can be understood by the BPRAM. One of the key factors is the formation of resonance states resulting from the hybridization between the valence (conduction) band and V$_\mathrm{Fe}$ (Fe$_\mathrm{V}$) antisite, as illustrated in Fig.~\ref{fig8}. Since a localized state of V$_\mathrm{Fe}$ is located just below $\varepsilon_\mathrm{F}$~\cite{Garmroudi2022,Bandaru2017}, the on-site Coulomb interaction on V$_\mathrm{Fe}$ easily induces the spin-split states (see Fig.~\ref{fig8}). This mechanism facilitates the emergence of local magnetic moments on antisites in Fe$_2$VAl, which is consistent with the field-induced magnetization in quenched Fe$_2$VAl~\cite{Garmroudi2022}. Furthermore, such induced local moments may play a crucial role in the magnetism of Heusler alloys, which in turn influences their thermoelectric properties~\cite{Tsuji2019}. 

Another crucial factor influencing the sign change of $S$ is the relative strength of scattering rates of valence and conduction electrons in semimetallic systems. This can be easily demonstrated using a simple toy model, in which the electronic states are described by the effective mass approximation near the band edges. In this toy model, the spectral conductivity $\sigma^\alpha(\varepsilon)$ for the valence and  bands (with spin index omitted) is proportional to $A_\alpha[(\tilde\varepsilon^2+\Gamma_\alpha^2)^{1/2}+|\tilde\varepsilon|^2]^{3/2}/\Gamma_\alpha$,~\cite{Matsuura2021} where $A_\alpha$ is a coefficient dependent on band degeneracy and effective mass,  and $\Gamma_\alpha$ represents the scattering rate. If $A_\mathrm{v}>A_\mathrm{c}$ and $\Gamma_\mathrm{v}>\Gamma_\mathrm{c}$ in semimetallic DOS similar to Fig.~\ref{fig6}(a), then $\sigma^\mathrm{v}(\varepsilon)<\sigma^\mathrm{c}(\varepsilon)$ near $\varepsilon_\mathrm{F}$, closely resembling the behavior in Fig.~\ref{fig6}(d). This relation naturally leads to negative $S$ according to Eq.~(\ref{L12}), despite the number of hole carriers exceeding that of electron carriers, as expected from the DOS under the condition of $A_\mathrm{v}>A_\mathrm{c}$. Our BPRAM of Fe$_2$VAl provides a microscopic understanding of this fact, attributing it to the hybridization between valence electrons of Fe3$d$ and V$_\mathrm{Fe}$ antisites. 

In summary, we have proposed a mechanism explaining the sign change of the Seebeck coefficient from positive to negative in Fe$_2$VAl due to the presence of antisite defects.  Based on the BPRAM, which describes hybridization between mobile carriers and randomly distributed impurities in the valence and conduction bands, we have shown that the scattering rate for hole carriers in the valence band is greater than that for electron carriers in the conduction band. The difference in scattering rates results in the negative sign of the Seebeck coefficient in spite of the density of states in the valence band exceeding that in the conduction band near the Fermi level. This mechanism could also be applied other Fe$_2$VAl-based compounds subjected to quenching procedures that induce antisite defects~\cite{Garmroudi2023b}.  Controlling scattering rates resulting from antisites will be a possible route to change thermoelectric properties in semimetallic systems without changing carrier concentration.

\begin{acknowledgment}
We thank T. Mori for helpful and informative discussions.  
\end{acknowledgment}

\end{document}